\documentstyle[12pt,psfig,epsf,epsfig,wrapfig,here]{article}
\input{epsf}                                               
                                                           
\begin{document}                                           
\begin{flushright}
\vglue-2.0cm
{\small\bf Invited talk, International Workshop on Electron Polarized Ion 
Collider, Bloomington, Indiana April 1999}
\end{flushright}

\begin{center}                                                           
{\Large{\bf {Diffractive Dissociation of High Momentum Pions}}}
{\large\footnote{Representing the E791 collaboration.
Supported in part by the Israel Science 
Foundation and the US-Israel Binational Science Foundation}}
\vglue0.5cm                          
{\large {\bf Daniel Ashery }}\\
\vglue0.5cm
{\em
School of Physics and Astronomy, Raymond and Beverly Sackler
Faculty of Exact Sciences Tel Aviv University, Israel}
\end{center} 
 
 
\begin{abstract}
The  diffractive dissociation of 500 GeV/c $\pi^-$ into di-jets is
described as a way to measure the momentum distribution of quarks in the
pion. The measurements of the pion diffractive dissociation
were carried out using data from Fermilab E791. Preliminary results show 
that the $|q\bar {q}\rangle $ Asymptotic
wave function which was developed using perturbative QCD methods describes
the data well for $Q^2 \sim 10 ~{\rm (GeV/c)^2}$. At these values signals of 
color transparency are expected and, indeed, observed through
the $A$-dependence of the yield of the diffractive di-jets.
\end{abstract}
        
\vglue-0.5cm
\section{Introduction}

One of the most sensitive tests of QCD is measurement of the internal
momentum distribution of quarks in hadrons. These distributions were 
calculated for the pion already about 20 years ago (section \ref{sec:thwf}) 
through calculations of the light-cone wave functions. However, 
there were no direct experimental measurements. Measurements of observables
which are related to these distributions, such as the pion electromagnetic form
factors, turned out to be rather insensitive to the light-cone wave functions.
The pion wave function is expanded in terms of Fock states:

\begin{equation}
\Psi = \alpha |q\bar {q}\rangle + \beta |q\bar {q}g\rangle +
           \gamma |q\bar {q}gg\rangle + ... .
\end{equation}
The first (valence) component will be dominant at large Q$^2$ as the other
terms are suppressd by powers of $m^2/Q^2$.

\section{The Pion Momentum Wave Function.}
\subsection{Theoretical predictions}
\label{sec:thwf}
Two functions have been proposed for the $|q\bar {q}\rangle$ configuration.
The Asymptotic function was calculated \cite{lb,er} using
perturbative QCD methods and is thus expected to be correct for
very large $Q^2$ ($Q^2 \rightarrow \infty$):

\begin{equation}
\phi_{as}(x) =\sqrt{3} x(1-x).
\end{equation}

\noindent
$x$ is the fraction of the longitudinal momentum of the pion
carried by a quark in the infinite momentum frame (see Fig. 
\ref{fig:pion_dis}).
Using QCD sum rules Chernyak and Zhitnitsky \cite{cz} proposed a function
that is expected to be correct for low Q$^2$:

\begin{equation}
\phi_{cz}(x) =5\sqrt{3} x(1-x)(1-2x)^2.
\end{equation}

\begin{wrapfigure}{r}{7cm}
\epsfig{figure=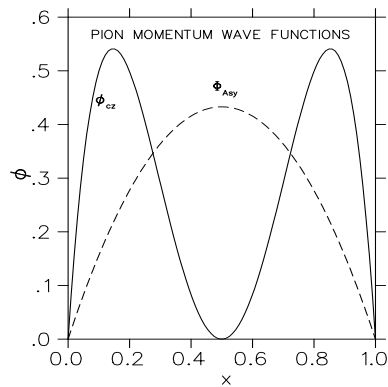,width=5cm,height=5cm}
\caption{Two predictions of the wave functions of the pion. Asymptotic
        function: dashed line, and CZ function: solid line.}
\label{fig:pionwf}
\end{wrapfigure}

The wave functions are plotted in Fig. \ref{fig:pionwf} and, as can be
seen, there is big difference between the two. Measurements of the
electromagnetic form factors of the pion were considered as the best
way to study these wave functions. A comprehensive summary of the status
of these measurements was recently published \cite{stesto}. Measurements
of the pion electric form factor suffer from two major drawbacks: measurements
done by elastically scatterred pions from atomic electrons measure
directly the  pion electric form factor but can be done only at very low
$Q^2$. The alternative method is measurements of $p(e,e' \pi^+)n$ in
the electron-pion quasi-free scattering kinematics. The relation between
the form factor and the longitudinal cross section is model dependent as
it includes the $p \rightarrow n\pi^+$ matrix element. Finally, the
form factor is related to the wave function through integral over
the wave function and the scattering matrix element reducing the sensitivity
to the wave function. Indeed, as shown in \cite{stesto} the two wave functions
can be made to agree with the experimental data. Similar situation
exists for inelastic form factors, decay modes of 
heavy mesons etc.  The problem is that
comparisons with these observables are not sensitive to details of
the wave function and thus cannot provide critical tests of the
$x$-dependence of the functions. An open question is also what can be
considered as high enough $Q^2$ to qualify for perturbative QCD
calculations and what is a low enough value to qualify for a treatement
based on QCD sum rules. An experimental study that provides information
about the momentum distribution of the $|q\bar {q}\rangle$ in the pion will
be described.


\subsection{Pion Diffractive Dissociation and Momentum Wave Function.}
\label{sec:pdif}

\begin{wrapfigure}{r}{7cm}
\epsfig{figure=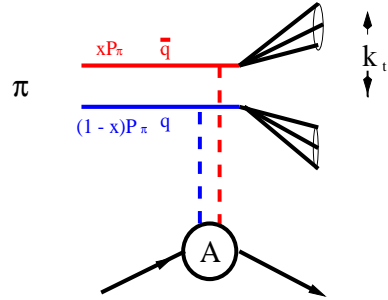,width=5cm,height=4cm}
\caption{measurement of the quark momentum wave function of the pion by
measuring longitudinal momentum of the two jets from diffractive 
dissociation.}
\label{fig:pion_dis}
\end{wrapfigure}

The concept of an experiment that maps the pion momentum wave
function is shown in Fig. \ref{fig:pion_dis}. A high energy pion
dissociates diffractively on a nuclear target imparting no energy 
to the target so that it does not break up. This is a coherent
process in which the quark and antiquark break apart and hadronize
into two jets. If in this fragmentation process the quark momentum
is transferred to the jet, measurement of the jet momentum gives the
quark (and antiquark) momentum. Then:

\begin{equation}
x_{measured} = \frac {p_{jet1}} {p_{jet1}+p_{jet2}}
\label{x_meas}
\end{equation}
By varying $k_t$ (Fig. \ref{fig:pion_dis}) the measurement can map 
the $x$ dependence for different  $|q\bar {q}\rangle$ sizes.

\begin{figure}[H]
\centerline{\epsfxsize=15cm \epsfbox{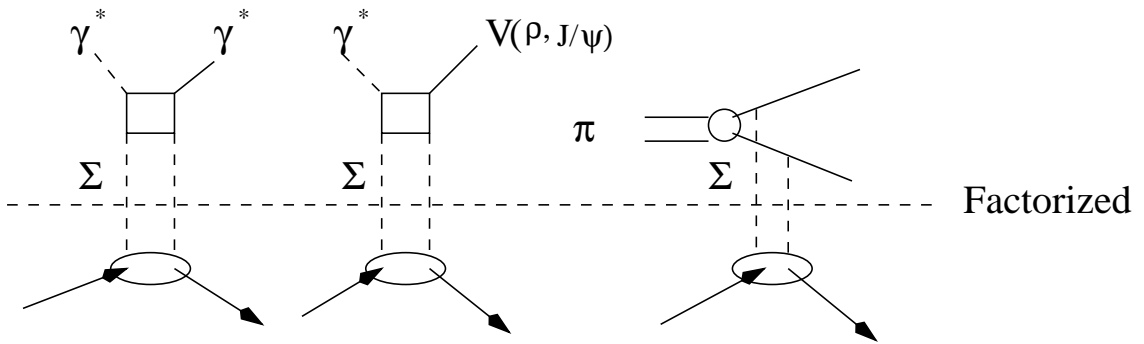}}
\caption{Factorization of inclusive DIS, vector meson production and
pion dissociation.}
\label{fig:fact}
\end{figure}

The diffractive dissociation of high momentum pions into two jets can be
described, like the inclusive Deep Inelastic Scattering
(DIS) and exclusive vector meson production
in DIS, by factoring the perturbative high momentum transfer process from 
the soft nonperturbative part \cite{fact}. This is described in Fig. 
\ref{fig:fact}.

This factorization allows the use of common parameters to describe the
three processes. The {\em Virtuality} of the process is given by $Q^2$,
the mass of the virtual photon, in inclusive and exclusive light
vector meson DIS. In exclusive DIS production of heavy vector
mesons this mass equals that of the produced meson ($J/\psi$). For
diffractive dissociation into two jets the mass of the di-jets is
playing this role. From simple kinematics and assumimg that the masses
of the jets are small compared with the mass of the di-jets, this can be
shown to equal $\frac{k_t^2}{x(1 - x)}$. $k_t$
is the transverse momentum of each jet. 

\newpage

\subsection{Fermilab E791 Experiment}

\begin{wrapfigure}{r}{8cm}
\epsfig{figure=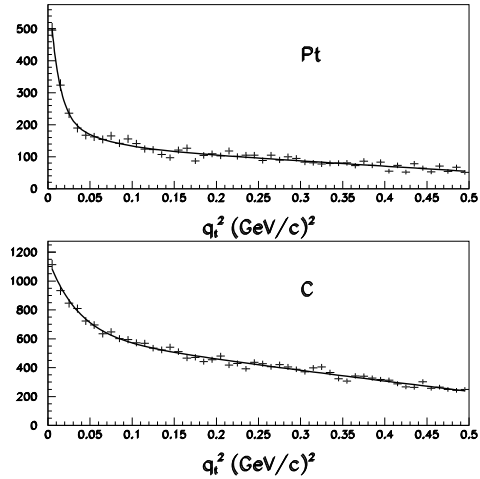,width=7cm,height=7cm}
\caption{$q_t^2$ distributions of di-jets from interaction of 500 GeV/c
$\pi^-$ with Carbon and Platinum targets.}
\label{data_diff}
\end{wrapfigure}

Fermilab experiment E791 \cite{e791}
recorded $2\times 10^{10}$ events from interactions of a 500 GeV/c 
$\pi^-$ beam with Carbon (C) and Platinum (Pt) targets.
The trigger included a loose requirement on transverse energy deposited
in the calorimeters.
Precision vertex and tracking information was provided by 23 silicon
microstrip detectors (6 upstream and 17 downstream of the targets),
ten proportional wire-chamber planes, and 35 drift-chamber planes. 
Momentum was measured using two dipole magnets. Two multicell, 
threshold \v{C}erenkov counters were used for $\pi$, $K$, and
$p$ identification. Only 1/3 of the E791 data was used for the analysis
presented here.\\

The data was analysed by selecting events in which 90\% of the beam
momentum was carried by charged particles. This reduced the effects
of the unobserved neutral particles. The selected events were
subjected to the JADE jet-finding algorithm \cite{jade}. The algorithm
uses a cut-off parameter ($m_{cut}$) that describes the mass of
individual jets. Its value was determined from Monte Carlo simulations 
to optimize di-jet detection. The di-jet invariant mass was calculated 
assuming that all the particles were pions. Only two-jet events were 
selected for further analysis. To insure clean selection of 
events, a minimum $k_t$ of 1.5 GeV/c was required. \\

Diffractive di-jets were identified through the $e^{-bq_t^2}$ dependence
of their yield ($q_t^2$ is the square of the transverse momentum transferred
to the nucleus and $b$ is propotional to the r.m.s. radius squared of the
nuclear target). Figure~\ref{data_diff} shows the $q_t^2$ distributions
of di-jet events from platinum (top) and carbon (bottom). The different 
slopes at low $q_t^2$ in the coherent region reflect the different 
nuclear radii. Events in this region come from diffractive dissociation
of the pion with the nucleus retaining its characteristic shape and
remaining intact.


\subsection{Results}

The basic assumption that the momentum carried by the dissociating
$q \bar q$ is transferred to the di-jets, was examined by Monte Carlo
(MC) simulation. Two samples of MC with 6 GeV/c$^2$ mass di-jets were generated 
with different $x$ dependences at the quark level. One sample simulated 
the Asymptotic wave function and the other the Chernyak Zhitnitski function.
The two samples were allowed to hadronize through the LUND PYTHIA-JETSET
\cite{mc} simulation and then passed through simulation of the experimental 
apparatus. 

\begin{wrapfigure}{r}{8cm}
\epsfig{figure=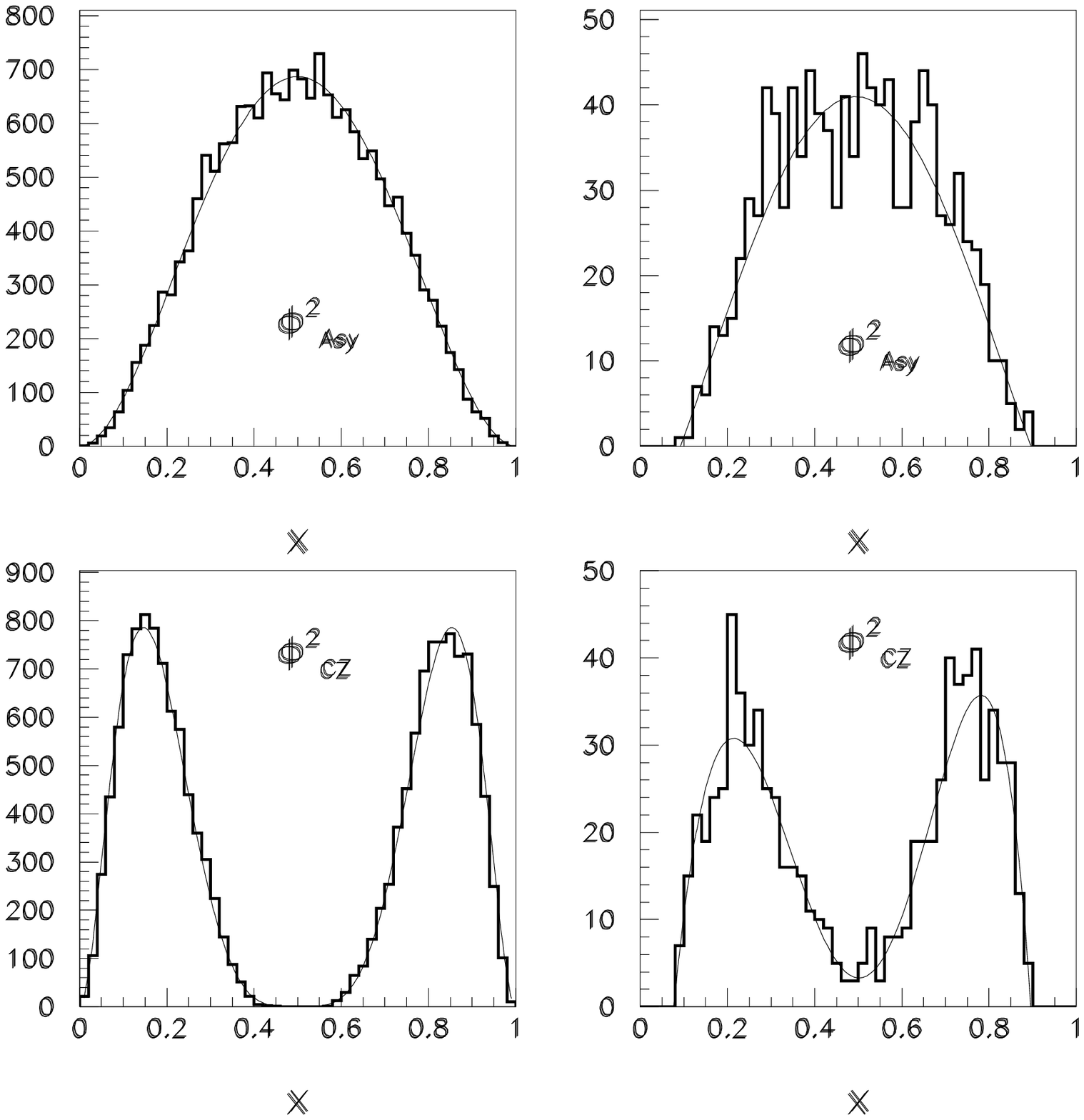,width=8cm,height=8cm}
\caption{Monte Carlo simulations of the two wave functions at the quark 
level (left) and of the reconstructed distributions of di-jets as detected
 (right).}
\label{fig:x_mc}
\end{wrapfigure}

In Fig. \ref{fig:x_mc} the initial distributions at the
quark level are compared with the final distributions of the detected 
di-jets, including distortions in the hadronization process and 
influence due to experimental acceptance. As can be seen the qualitative
features of the two distributions are retained.\\

\begin{figure}[H]
\centerline{\epsfxsize=6.2cm \epsfbox{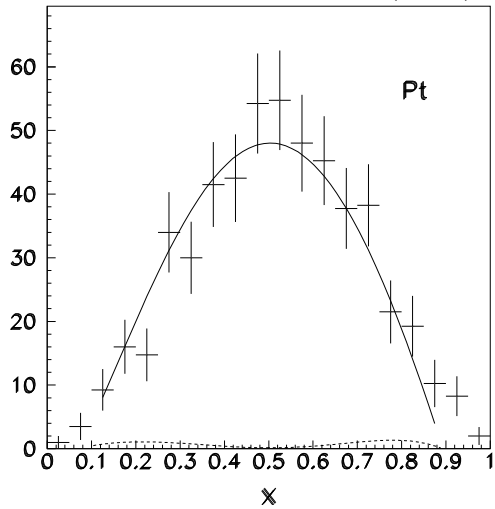}}
\vglue -1.0cm
\caption{The preliminary $x$ distribution of the diffractive di-jets fron the
         platinum target. The line is a fit to a wave function with 
         about 90$\%$ asymptotic and 10$\%$ CZ. The CZ contribution is
         shown with a dashed line.}
\label{xdatadif}
\end{figure}

For events in the coherent region described in the previous section the 
value of $x$ was computed from the measured longitudinal momentum of each 
jet (eq. \ref{x_meas}). The resulting $x$ distribution 
is shown in Fig. \ref{xdatadif}.
Superimposed on the data is a fit to a combination of the two simulated
final  di-jet distributions of Fig. \ref{fig:x_mc}.
\vglue -1.0cm

The preliminary result
under the analysis conditions as described is that the data can be
fitted well with $\geq$ 90\% of the Asymptotic wave function 
and $\leq$ 10\% of the CZ wave function. The
requirement that  $k_t > $ 1.5 GeV/c can be translated to $Q^2 \sim 10 
~{\rm (GeV/c)^2}$ (section \ref{sec:pdif}). This shows that for these $Q^2$ 
values the perturbative QCD approach that led to construction of the asymptotic 
wave function is reasonable.


\section{Pion Diffractive Dissociation and Color Transparency}

\subsection{Theoretical Predictions}

Diffractive dissociation of pions to di-jets with $k_t ~>$ 1.5 GeV/c
($Q^2 \sim 10 ~{\rm (GeV/c)^2}$) can be
associated with a $|q\bar {q}\rangle$ system with size 
$<r>_{|q\bar {q}\rangle} ~\sim ~1/Q ~\leq 0.1$ fm. Such a small size
configuration is expected to exhibit the effect of color transparency,
namely a reduced interaction with the nuclear medium. The
condition for observation of the effect is that the system does not
expand while traversing the nucleus. The distance that the system passes
before it decays and expands (the ``Coherence Length") is estimated
by  $l_c ~\sim~\frac{2p_{lab}}{M^2-m_{\pi}^2}$ where M is the di-jet mass.
It is required that $l_c ~~> ~~R_A $. In E791 $2p_{lab}$ ~= ~1000 GeV/c.
Using M $\sim$ 6 $~{\rm GeV/c^2}$ the result is ~$l_c ~\sim$ ~20 ~fm. which is 
larger than all nuclear radii.\\

Bertch et al. \cite{bbgg} proposed that when high momentum pions hit
a nuclear target the small  $|q\bar {q}\rangle$ component will be filtered
by the nucleus and materialize as diffractive di-jets. 
Frankfurt et al. \cite{fms} proposed that  diffractive dissociation of 
pions to high $k_t$ di-jets {\em selects} the small size $|q\bar {q}\rangle$ 
Fock state with the size controlled by  $k_t$. They calculate the cross 
section for this  diffractive dissociation of pions on nuclear 
targets with the striking result that for $t$ = 0 and $k_t > $ 1.5 GeV/c

\begin{equation}
\frac{d^4 \sigma_A}{dx dM^2_J d^2 q_t} =
  \frac {d^4 \sigma_N}{dx dM^2_J d^2 q_t}  A^2 ~~~~~(t = 0)
\end{equation}

Namely that the differential  cross section at $t$ = 0 is proportional to 
$A^2$ which is a very unusual dependence. The $t$-dependence 
for t $\neq$ 0 is given by the nuclear form factor:
$|F_A(t)|^2 \sim e^{bt}$,  ~~~$b = \frac{<R^2>}{3}$.

\subsection{Expected Diffractive Di-Jet A-dependence }

Realistically, it is impossible to measure the cross section at $t$ = 0
for two reasons: there is always a minimal $t$ transferred to the target
as a result of excitation of the pion to a heavy mass, and the
statistical and systematic uncertainties near $t$ = 0 make it difficult
to obtain precise results. \\

The minimal $t$ transfer can be estimated from simple kinematics. The result
for a pion being excited from $m_{\pi}$ to the di-jet mass  $M_{jets}$
is a longitudinal momentum transfer of:

\begin{equation}
q_z = \frac { M_{jets}^2 - m_{\pi}^2 - 2E_1\frac {t}{2 M_A} - t}{ 2p_1}\\
\end{equation}
with E$_1$ the bombarding energy. The energy transferred to the nucleus is:

\begin{eqnarray}
 q_0 &= & - \frac {t}{2 M_A}.
\end{eqnarray}

For a numerical estimate we use:

\begin{eqnarray}
-t<1/b=\frac{3}{r^2 \;{\rm fm}^2}
{\rm ~~Carbon} & \approx & 2.0 \times  10^{-2} \; ({\rm GeV/c})^2 \\\nonumber
{\rm ~~~~~Platinum} & \approx & 4.3 \times 10^{-3} \; ({\rm GeV/c})^2
\end{eqnarray}
$t \times E_1/M_A$ $\approx$ 0.83 $({\rm GeV/c})^2$ (carbon) and 0.011
$({\rm GeV/c})^2$ (platinum), $ m_{\pi}^2=0.02$
We can therefore neglect all the terms except for $M_{jets}^2$ and write

\begin{equation}
q_z \simeq \frac { M_{jets}^2}{ 2p_1}
\end{equation}

The minimal longitudinal momentum-square transferred to the nucleus 
as a consequence of exciting the pion to $M_{jets}$ is
$q_z^2 = -t_{min}$. With $M_{jets} \sim 6 ~{\rm GeV/c^2}$ we get:

\begin{equation}
-t_{min} = q_z^2 \simeq (\frac{M^2_{jets}}{2p_{\pi}})^2
            \simeq  (\frac{36}{1000})^2 = 0.0013 \;\; (GeV/c)^2.
\end{equation}

\noindent
While this value is small, it is not negligible compared with the $t$
values for Pt. In order to avoid this problem and to increase
the precision of the measurement we integrate the data in the 
diffractive region. This modifies the expected $A$ dependence. We
first write:

\begin{equation}
\frac {d\sigma}{dt} = \sigma_N A^2 e^{bt} =
                               \sigma_N A^2 e^{-b(q_0^2+q_z^2+q_t^2)}
\end{equation}
with $q_t^2$ the transferred transverse-momentum  squared. From the
numerical estimates we see that $q_0^2$ can be neglected so that
$~t = t_{min} - q^2_t$ and $d\sigma / dt  ~~\rightarrow ~~d\sigma / 
dq_t^2 $. Integration of the diffractive region (using $R=R_0 A^{1/3}$)
yields:

\begin{equation}
\int (d\sigma / dq_t^2) dq_t^2 = \sigma_N A^2 e^{bt_{min}} e^{-bq_t^2}
                            \times (1/b) =
\sigma_N A^2 e^{bt_{min}} e^{-bq_t^2} \times \frac{3}{R_0^2 A^{2/3}}
\end{equation}
and the expected general $A$ dependence is: 
$\frac{d^2 \sigma}{dx dM^2} \propto A^{4/3}$

In the particular case of E791, using the numerical values of the nuclear
radii of C and Pt:

\begin{equation}
\left(\frac{R_{Pt}}{R_C}\right)^2 = \left(\frac{5.27}{2.44}\right)^2
= 4.665 = \left(\frac{195}{12}\right)^x; ~~~x = 0.55.
\end{equation}
so that we will expect $\frac{d^2 \sigma}{dx dM^2} \propto A^{1.45}$.

\subsection{Analysis and Results}

\begin{figure}[H]
\centerline{\epsfxsize=13.0cm \epsfbox{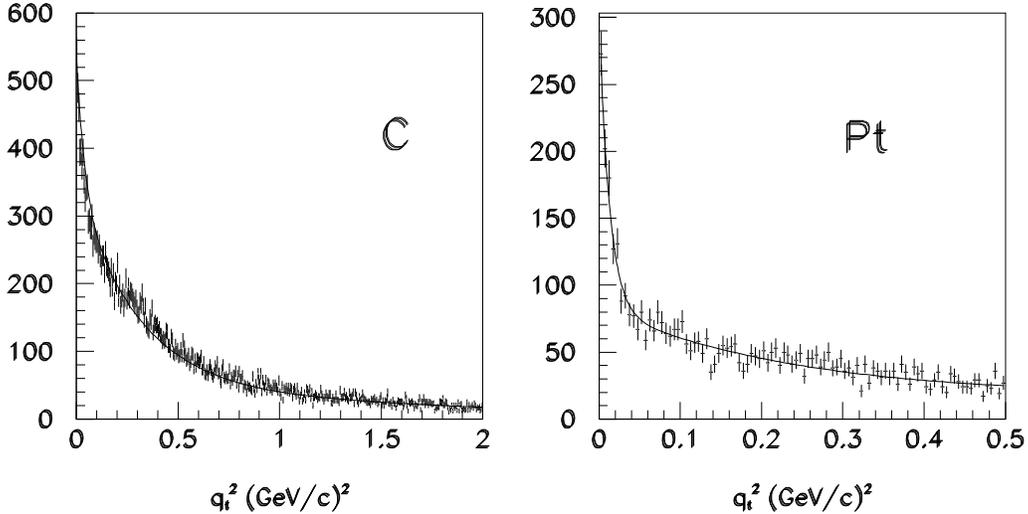}}
\vglue -8.0cm
\caption{$q_t^2$ distributions of di-jets for C and Pt targets. The
lines are fits of the MC simulations to the data.}
\label{difr_a}
\end{figure}

To obtain the $A$ dependence of the diffractive yield from the
experimental results of Fig. \ref{data_diff} we must realize that these
results are a combination of three elements: coherent nuclear diffraction,
incoherent nuclear diffraction (interaction with individual nucleons)
and smearing of both components by the experimental apparatus. In order
to extract the coherent nuclear diffractive yield we simulate the
coherent and incoherent $q_t^2$ distributions of the dissociation by 
assigning them the nuclear and the nucleon form factors, respectively. 
The simulated distributions are passed through the simulated experimental
system so that the individual reconstructed distributions are obtained. At
this stage a combination of the reconstructed distributions of the Pt 
nucleus and that of the nucleon is fitted to the data from the Pt target
giving the ratio of coherent/incoherent yields and an overall
normalization factor between the data and the simulation. The same is
done for the C target. Results of these fits are shown in 
Fig. \ref{difr_a}.

The final results are obtained by integrating over the diffractive 
regions (of Pt and C) in the generated MC  multiplied by the total 
normalization factor between MC and data. The analysis was done
separately in two $k_t$ ranges since, as predicted by \cite{fms} there
can be some dependence of $\alpha$ in $\sigma \propto A^{\alpha}$ on $k_t$.
The PRELIMINARY result, compared with those expected according to
\cite{fms}, labelled $\alpha$(CT) are:

\vspace{0.5cm}

\begin{tabbing}
12345678901234567812 \=
12345678901234567812 \=
1234567890123456 \=
\kill

\underline{k$_t$ range} \> ~~~~~~~\underline{$\alpha$} \> 
\underline{$\alpha$(CT)~} \\
\\

1.5 $ < ~k_t <$ 2.0 \> 1.61 $\pm$ 0.08 \> 1.45 \\
\\
2.0 $ < ~k_t$ \> 1.65 $\pm$ 0.09 \> 1.60 \\

\end{tabbing}

\noindent
which are close to the Color Transparency expectations.

\vglue -2.0cm

\section{Summary}

          The following observations have been made:\\

\begin{itemize}
\item{Diffractive dissociation of hadrons into di-jets can be used to study 
the hadron's internal quark structure.}

\item{The momentum wave function of the $|q\bar{q}\rangle$ in the pion is
described well at Q$^2 ~\sim ~10 ~{\rm (GeV/c)^2}$ ($k_t >$ 1.5 GeV/c) by the 
Asymptotic wave function.}

\item{Using data from Pt and C targets and assuming $\sigma \propto A^{\alpha}$
the cross section for diffractive dissociation of pions to di-jets is found 
to be proportional to $\sim ~A^{1.61 \pm 0.08}$ for 1.5 $ < ~k_t <$ 2.0 GeV/c
and $\sim ~A^{1.65 \pm 0.09}$ for 2.0 GeV/c $ < ~k_t $ .
These results are consistent 
with predictions based on the color transparency effect.}

\end{itemize}

\end{document}